# Spatio-Temporal Amplitude-and-phase Reconstruction by Fourier-transform of Interference Spectra of High-complex-beams

**Benjamín Alonso,[1,*] Íñigo J. Sola,[1] Óscar Varela,[1] Juan Hernández-Toro,[1] Cruz Méndez,[2] Julio San Román,[1] Amelle Zaïr,[1] and Luis Roso[2]**

[1]*Universidad de Salamanca, Área de Óptica. Departamento de Física Aplicada, E-37008 Salamanca, Spain.*
[2]*Centro de Láseres Pulsados Ultracortos Ultraintensos, CLPU, E-37008 Salamanca, Spain.*
[*]*Corresponding author: b.alonso@usal.es*

We propose a novel method to reconstruct the spatio-temporal amplitude and phase of the electric field of ultrashort laser pulses using spatially-resolved spectral interferometry. This method is based on a fiber-optic coupler interferometer that has certain advantages in comparison with standard interferometer systems, such as it being alignment-free and selection of the reference beam at a single point. Our technique, which we refer to as STARFISH, offers compactness and simplicity. We report its application to the experimental characterization of chirped pulses and to spatio-temporal reconstructions of a convergent beam as well as plane-plane and spherical-plane waves interferences, which we check with our simulations.

## 1. INTRODUCTION

Laser pulse characterization is a fundamental issue for the extensive community of laser users. Knowledge of the structure of laser beams before they are used in certain experiments or applications is essential, as is study of the pulse after it has undergone a given process. However, the complexity of the electric field requires certain specific approaches when addressing it. The electric field of the light is in general a vector that depends on time and three spatial variables (two transverse directions and the longitudinal direction of propagation). We shall neglect the vectorial character of the electric field because here we are dealing with linearly polarized pulses.

The temporal variation and the spatial profile of ultra-short laser pulses are often characterized separately. Thus, the temporal characterization of pulses has a fairly long history. Pulse autocorrelation can afford an idea of the laser pulse duration and form [1]. Several robust and reliable techniques are now well established for amplitude and phase retrieval, such as Frequency Resolved Optical Gating (FROG) [2], Spectral Phase Interferometry for Direct Electric-field Reconstruction (SPIDER) [3], spectral interferometry (SI) [4,5] and its combination with FROG, known as Temporal Analysis by Dispersing a Pair of Light Electric-fields (TADPOLE) [6]. Additionally, the spatial profile can be measured with a standard detector (a CCD), but the spatial phase (i.e. the wave-front) can also be measured with different techniques, e.g. Hartmann-Shack [7].

One approach to the study of spatio-temporal evolution consists in measuring the temporal pulse profile at different spatial positions with SPIDER [8,9], and the GRENOUILLE (single-shot FROG) can also be used to measure pulse-front tilt [10]. Nevertheless, in spite of being very useful these techniques cannot measure the full spatio-temporal information and preserve the coupling. The aim of our work is to develop a simple and robust system capable of reconstructing the spatio-temporal amplitude and phase of laser pulses at a fixed propagation distance. Our underlying motivation is the broad variety of application fields of spatio-temporal characterization, such as the study of optical aberrations [11,12] or nonlinear propagation [13-15].

Initial schemes for this purpose were based on spatially-resolved SI and did not characterize the reference beam, thus only measuring phase differences and being unable to perform complete spatio-temporal reconstructions [11,15]. Spatially resolved SI consists in measuring the spectral interferences of a test and a reference beam across the spatial profile. If the spatio-spectral phase of the reference beam is known, the spatio-temporal coupled amplitude and phase of the test pulse can be retrieved. In the scheme of Diddams et al. [16], a Mach-Zehnder interferometer was used. Those authors filtered the reference beam spatially and measured it in a single position, assuming a constant spectral phase. In a previous work, we implemented the same scheme and found that the spatial cleaning of the reference beam is not easy and does not ensure a perfectly homogeneous reference, especially when filtering complex pulses [17].

The amplitude and phase reconstruction of spatio-temporal coupling was studied by Rivet et al. [18], using a combination of Hartmann-Shack and FROG or SPIDER. This scheme, known as Shackled-FROG, has recently been demonstrated by Rubino et al. [19]. A robust technique used by Dorrer et al. consisted of two-dimensional (spatial and spectral) shearing interferometry [20]. More recently, a holographic method reported by Gabolde et al. (STRIPED FISH) [21] has demonstrated the ability to measure three-dimensional (x,y,t) electric fields in single-shot. To study non-linear propagation, Trull et al. developed a technique that measures the spatially resolved temporal cross-correlation [13], measuring the sum frequency with a short probe. This technique provides an image of the spatio-temporal intensity (but not the phase), which is valuable information that has been used for measurements of X-waves [14].

Another recent technique is based on Spatially Encoded Arrangement Temporal Analysis by Dispersing a Pair of Light Electric-fields (SEA TADPOLE) [22] scanning the test beam as proposed by Bowlan et al. [12]. The SEA TADPOLE technique measures the spectrally resolved spatial interferences of two non-delayed crossed beams. This idea had already been implemented by Meshulach et al. [23] in a primitive scheme involving crossing the beams directly, and has been adapted by Bowlan et al. by guiding a spatial selection of each beam (test and reference) with equal-length, single-mode optical fibers. The temporal profile of the test beam is recovered from the spatio-spectral trace, hence with the advantage of using the full spectrometer resolution. The extension to spatio-temporal characterization merely consists in scanning the test beam profile with the fiber.

In this contribution, we report a novel scheme for spatially resolved SI based on a fiber-optic coupler interferometer that has certain advantages in comparison with standard interferometers. We refer to it as Spatio-Temporal Amplitude-and-phase Reconstruction by Fourier-transform of Interference Spectra of High-complex-beams (STARFISH).

Our system bears some similarities to SEA TADPOLE since the test and reference beam are collected with optical fiber inputs: both systems are free of alignment and use a single reference pulse. The main differences are that our system is based on SI instead of on spatial interferences (SEA) and that STARFISH only uses a fiber coupler and a standard spectrometer. This makes our system very simple and robust in experimental terms, and its implementation or upgrading in a laboratory in a plug-and-play scheme is easier (simply plugging the coupler to the spectrometer). Moreover, STARFISH measures a single spectrum for each spatial point (instead of a spatio-spectral trace), thereby reducing data processing, which would be more interesting -for example- when measuring many spatial points at different propagation distances. Despite this, the use of standard SI instead of spatial arrangements (SEA) involves a loss of spectrometer resolution, thus limiting the pulse length capable of being measured with STARFISH in comparison with SEA TADPOLE [22] and SEA SPIDER [24], although we shall show that this is not a problem.

## 2. SPATIALLY-RESOLVED SPECTRAL INTERFEROMETRY

### A. Spectral Interferometry

In spectral interferometry, two beams (test and reference) are delayed with respect to each other by a time $\tau$, and propagate collinearly. The resulting spectrum of both beams is the sum of the spectra plus an interference term containing the information of the phase difference between the beams, as seen in Eq. 1. The interference fringes have a period given by the inverse of the delay. For the sign, we chose the criterion that the reference always goes before the test and the delay is positive. The resulting spectrum is

$$S(\omega) = S_{test}(\omega) + S_{ref}(\omega) \\ + 2\sqrt{S_{test}(\omega)S_{ref}(\omega)} \cos[\phi_{test}(\omega) - \phi_{ref}(\omega) - \omega\tau] \quad (1)$$

The fringe-inversion technique, the so-called Fourier Transform Spectral Interferometry (FTSI) [25], can be applied to retrieve the phase information. The inverse Fourier transform affords three peaks in the temporal domain: one centered at $t=0$, coming from the continuum spectra (test and reference spectra), and two side peaks centered at $t=\pm\tau$, corresponding to the interference term. The continuum contribution ($t=0$) can be depleted by subtracting the test and reference spectra from the interference spectrum. One side-peak is filtered and returned to the spectral domain by direct Fourier transform, from where the phase difference between the test and the reference beam is extracted. The time delay, $\tau$, must be high enough to prevent central- and side-peak overlap, but small enough to allow the fringes to be resolved by the spectrometer. The reference spectrum must at least comprise the whole test spectrum and the spectral amplitude should be comparable to have well contrasted fringes. If the reference phase is known, the test spectral phase can be calculated and together with the test spectrum, the beam can be fully characterized in the temporal domain simply by applying a Fourier transform. The delay is calculated from the side-peak position at a certain point and the term $\omega\tau$ is added to the retrieved phase.

The extension of SI to spatio-temporal characterization in standard interferometer schemes is achieved by a spatial reference that is delayed with regards to the test beam, as shown in Fig. 1. The reference beam must be spatially homogeneous (a flat wave-front and a spectral phase independent of the transverse position) because in general it can only be characterized at a single point. In this scheme, the test beam is referenced at each spatial position by a known pulse (thanks to the homogeneous spatio-temporal reference beam) and hence the connection between different spatial points in the test profile can be obtained. Accordingly, the experimental measurements consist in measuring the spectral interferences as a function of the scanned transverse position of the beam profile; that is, spatio-spectral traces depending on the position and the wavelength. In Eq. 1, this involves spectral amplitude and phase as a function of the wavelength and the transverse position. The spatially resolved spectrum (spectral trace) of the test arm is also measured, which, together with the spectral phase retrieved by the FTSI, allows the test beam to be characterized.

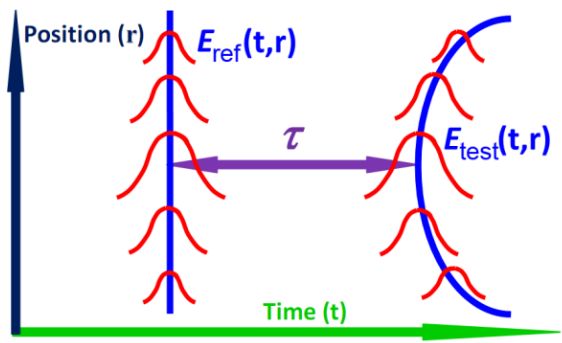

Fig. 1. (Color online) Scheme of spatio-temporal reference for spatially resolved spectral interferometry, consisting in using a homogeneous flat reference beam delayed with respect to the test beam and scanning the position (transverse), measuring their respective spectral interferences. Thus, each spatial position is referenced by a known pulse.

### B. Experimental setup for STARFISH: the fiber-optic coupler

To avoid the complications of standard interferometer-based systems (homogeneous reference, precise alignment…), here we propose an interferometer based on a fiber-optic coupler for SI (STARFISH). The fiber coupler must be single-mode to avoid different mode dispersions. The fiber coupler was designed to work within a broadband spectral region ranging from 680 to 900 nm, allowing the characterization of ultrashort pulses. The configuration of the fiber coupler comprises two input ports and a common output port. The reference and the test beam enter through each input port, are coupled in the transition, and exit the fiber, delayed, through the same port. In the experimental scheme (Fig. 2), the unknown beam is split, each replica being sent to each fiber arm. The test arm fiber input has a motorized stage for the spatial scan (transverse), whereas the position of the fiber for the reference arm controls the delay between the pulses (longitudinal). The reference arm selects a suitable spatial position containing all spectral components that the test beam has at any position. This reference is characterized temporally with the FROG or SPIDER techniques. Since the reference beam is not spatially scanned with the fiber input, only one point of the reference profile is selected with the fiber, and this allows a constant reference to be used instead of a possibly inhomogeneous reference profile. Thus, here we did not implement any spatial filtering because it was not necessary. Following the scheme shown in Fig. 1, our proposal consisted in referencing all the positions of the test beam by a single spatial point of the reference beam.

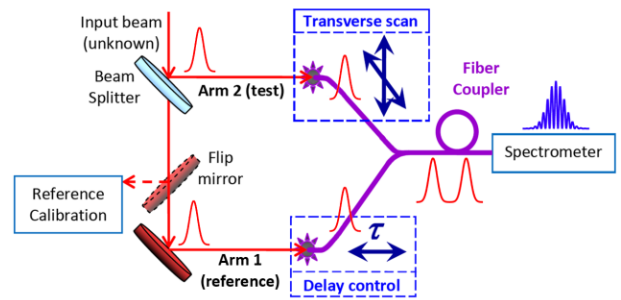

Fig. 2. (Color online) Set-up based on the fiber-optic coupler interferometer for spatially resolved spectral interferometry. The longitudinal position of one fiber arm controls the relative delay between reference and test beams. The test beam is scanned transversely (spatial) with its corresponding input fiber arm.

The arms of the fiber coupler should be equal length (this does not exactly occur in reality) in order to avoid introducing different dispersions on the beams. Nevertheless, we consistently calibrated the global phase difference of the fiber coupler (due to slight differences in the length of the arms of the fiber coupler) and the beam splitter. This calibration was accomplished by taking a single spectral interferometry measurement using the same input beam in both arms. It was then taken into account in the reconstruction algorithm to retrieve the correct spectral phase.

The main advantages of this system are its simplicity, compactness, the fact that alignment is not necessary, the ease with which it can be upgraded simply by connecting the coupler to a different spectrometer, and the fact that a constant reference is used (we selected the broadest spectrum region). This means that we do not need a spatial filter for the reference and approximations concerning reference homogeneity are required. In comparison with standard interferometers, we do not need a spatially homogeneous reference beam with at least the same spatial section and spectrum as the test beam. Moreover, the spatial resolution is given by the mode-field diameter of the single-mode fiber: in our case $4\mu m$.

## 3. EXPERIMENTAL MEASUREMENTS

### A. Introduction

The experiments were carried out using two different terawatt-class Ti:Sapphire CPA laser systems (both at a 10 Hz repetition rate). The first system (Spectra Physics, Inc.) delivers laser pulses of 120 fs (Fourier limit) with its spectrum centered at 795 nm. The second system (Amplitude Technologies) provides 35-fs pulses centered at 805 nm. We worked with two different lasers to test STARFISH with pulses of different durations and bandwidths. For the temporal characterization of the reference beam we used GRENOUILLE (single shot FROG, Swamp Optics) and SPIDER (APE GmbH) devices, whereas for the spectra we used a commercial spectrometer (AVANTES, Inc.). Depending on the duration of the pulses, we characterized them with the SPIDER (35-fs pulses), where there is no ambiguity in the time direction, or with the GRENOUILLE (120-fs pulses). In the case of the GRENOUILLE device, we identified the temporal direction by performing a second measurement with additional known dispersion, as is usually done when using this apparatus. We observed that the GRENOUILLE spatial homogeneity requirements were fulfilled for the 120-fs laser by measuring the profile with a CCD. The spatial scan was performed with a motorized micrometric stage at the same time as the spectrum was acquired.

We first tested our system for one-dimensional SI, checking the reconstruction of laser pulses in comparison with the SPIDER and GRENOUILLE characterizations. Among others, we performed a test in which the delay was varied from -5 ps to +5ps in 50-fs steps, and we observed that the pulse intensity and phase were reconstructed independently of the delay over a wide range: up to several picoseconds. We also checked our reconstruction algorithm with simulations involving the characterization of complex pulses.

In general, interferometry is affected by small fluctuations due to system instabilities and, in particular, our fiber coupler approach was indeed expected to exhibit a phase drift. The consequence is a variation in the relative phase term between the two interferometer arms. Since this variation is almost independent of the wavelength, this means a loss of the constant zero-order relative phase of the pulses, thus preventing precise knowledge of the pulse wave-front and introducing a small error in the pulse front. For our purposes, this was not a problem and indeed there are techniques for overcoming this drawback and retrieving the wave-front from this kind of measurement [26]. We studied the stability of the interferences for single-shot measurements (acquired continuously without average), tracking the zero-order phase, the full spectral phase, the delay and the time-width of the reconstruction. We measured a zero-order phase drift in the interferometer of 1.4 rad (peak to peak) during the time usually taken for a measurement to be made (about 1 minute). The phase drift (0.45π) slightly affected the pulse front, the corresponding temporal shift thus being limited to 0.60 fs (0.225T, with T the laser period). For the delay, we calculated a standard deviation of 0.40 fs and for the time-width of the measured pulse 0.09 fs, whereas the whole spectral phase was very stable between shots (except for the effect of the zero-order phase drift). We also studied stability for repeated multi-shot averaging measurements, for which we obtained blurry and reduced contrast interferences due to the shift of the fringes. As a result, more stable delays but poorer reconstructions were obtained due to incorrect phase retrieval.

### B. Linear chirp experiments

In order to explore the limitations of our setup for SI, we performed an experiment to measure the linear chirp. The test pulse was chirped through two passes in a diffraction-grating pair compressor using the 35-fs laser. We negatively chirped the pulse with Group Delay Dispersion (GDD) varying from $-7000 fs^2$ to $-1000 fs^2$ because these were the compressor limits for our setup. The linear chirp stretches the test pulse and this implies that the side-peaks in time of the Fourier-transform of the interferences broaden and decrease in amplitude. In our GDD scan, we varied the grating-distance, $L$, and hence the GDD calculated as in [27] is

$$GDD(L) \simeq -\frac{\lambda^3}{\pi c^2}\frac{L}{d^2 \cos^2 \theta_m}, \quad (2)$$

where $\lambda$ is the central wavelength, $c$ is the speed of light, $1/d = 300\, gr/mm$ gives the groove density, and $\theta_m$ is the output angle calculated from the grating equation $\sin\theta_m - \sin\theta_0 = m\lambda/d$ (for the first order $m=1$ and the incidence angle in the grating $\theta_0 = 15^o$). The GDD is linearly dependent on the grating distance and from Eq. 2 we calculated the estimated slope $GDD(L)/L = -212.8 fs^2/mm$. We measured the chirped pulses using the fiber coupler interferometer at 81 grating-distances and reconstructed them with FTSI. Thus, we obtained the spectral phase and calculated the experimental GDD from a quadratic fit, as shown in Fig. 3a, which corresponds to $GDD = -5200 fs^2$. In Fig. 3b the GDD is represented as a function of the grating distance. The linear regression of these data afforded a slope of $-210.2 fs^2/mm$, in very good agreement with the estimated value. Extrapolation of the fit to zero distance gives an acceptable deviation, $GDD(L=0) = -28.7 fs^2$, and the correlation coefficient was $R = 0.99984$, revealing the good fit to the data. We also checked that the possible third-order dispersion (TOD) was completely negligible as compared to the GDD. Finally, we studied the instantaneous wavelength (as a function of the time) of the pulses, calculated from the electric field phase. In Fig. 3c, using a false color scale we plot the instantaneous wavelength of the pulses as a function of the grating distance. We have cropped the plot for the decrease in pulse intensity greater than three orders of magnitude (shown in white). In this figure, we show the linear dependence on time of the instantaneous wavelength, explaining the pulse stretching. In Fig. 3d, We represent the temporal reconstruction and instantaneous wavelength of the pulse corresponding to $GDD = -5200 fs^2$. We measured chirped pulses as long as 1.3 ps ($1/e^2$ width, decrease in intensity to 13.5%) for the highest GDD. We also explored the intensity profile caused by the GDD (available at Media 1) and found that for the lowest chirps pulse splitting occurred, due to the spectrum profile, but not to TOD (negligible), which we also observed with the SPIDER measurements of the test pulse.

To further complete the results, we performed simulations using the experimental spectrum (bandwidth of 70-nm) and chirping the pulse with negative and positive GDD up to $40000 fs^2$, always imposing our spectrometer resolution, and found that we could measure 7-ps-long pulses ($1/e^2$ width). We also carried out simulations with the 120-fs laser spectrum with GDD from $-80000 fs^2$ to $80000 fs^2$ and retrieved the input GDD and reconstructed 4-ps pulses ($1/e^2$ width).

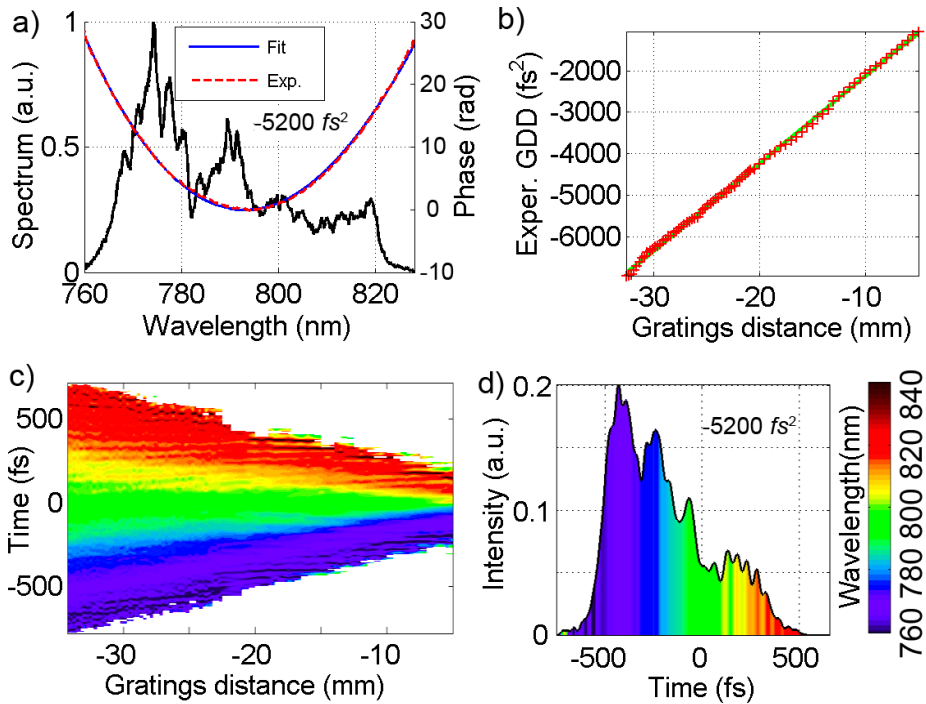

Fig. 3. (Color online) Experimental spectrum and phase of a negatively chirped pulse (a). Experimental scan on negative linear chirp: GDD retrieved from FTSI (b) and instantaneous wavelength of the chirped pulses (c) as a function of the grating-distance. Temporal intensity and instantaneous wavelength of the chirped pulse (d). The intensity profile and instantaneous wavelength variation with the GDD can be seen in a video available at Media 1.

One of the main advantages of the technique is its simplicity and robustness. These characteristics allow the reconstruction system to be adapted immediately to spectrometers or monochromators with much more resolution simply by plugging the fiber coupler to the input port of the device (common in most systems). Even though the whole spectral resolution of the spectrometer is not used, it is very easy to upgrade STARFISH with commercial devices, with resolutions of around 0.02 nm for portable and small spectrometers in the visible and the infrared (capable of measuring Fourier-transform-limited narrowband pulses of 5-ps FWHM, and even longer in case of broadband chirped pulses); around 0.004 nm for Optical Spectrum Analyzers (OSA) and monochromators in the visible and the infrared (compatible with pulses of 25 ps FWHM), and even below 100 fm in the mid-infrared range, such as in the BOSA High-Resolution Optical Spectrum Analyzer (allowing unchirped pulses of 1 ns FWHM to be reconstructed).

**C. Convergent wave**

The first spatio-temporal result reported here corresponds to a convergent wave using the 35-fs pulse duration laser focused by a 50-cm focal length lens (Fig. 4). The test beam (unknown beam) is scanned transversely 31 cm after the lens; that is, before the focus. In experiments with 35-fs pulses, the reference was calibrated with the SPIDER device. The delay between the reference and the test beam was 550 fs. In Fig. 4a, we show the spectral interference trace of the reference and test beam as a function of the wavelength and the transverse position. We scanned 4 mm of the beam profile in $20\mu m$-steps (201 points). The evolution of the fringes with the position is quadratic, in agreement with the curvature of the wave-front and the pulse-front of the test beam. The spatio-temporal intensity reconstruction is shown in Fig. 4b, in which the convergence of the beam may be observed: the peripheral region of the beam arrives before the central region at a certain propagation distance. We fitted the retrieved pulse-front curvature of the beam (see fit in blue dashed line in the figure) and obtained a value of 18.6 cm for the radius of curvature, in agreement with the expected value of 19 cm, if Gaussian beam propagation is assumed.

Since we used a terawatt laser, the beam profile was inhomogeneous and the pulse energy fluctuated. To remove the energy instability, we have averaged the test beam spectrum taken at each point in the measurements presented in this work. We checked that the spectral phase retrieved was not affected by this instability. The intensity reconstruction showed in Fig. 4b exhibits spatial modulations that can be explained in terms of the spatial inhomogeneity of the beam. We checked that it was only due to the spatial profile by directly comparing the reconstruction with the Fourier-transform limit of the spatially resolved spectrum (where the FTSI cannot cause them). For further proof, we checked the reproducibility of the beam profile reconstruction, thus discarding SI or laser instability as being the origin of the inhomogeneity of the reconstructed profile.

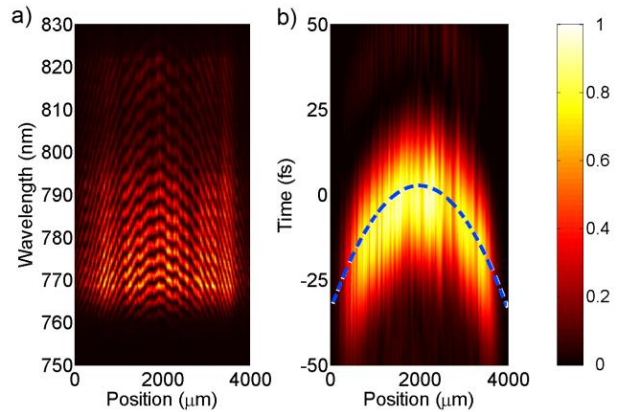

Fig. 4. (Color online) Spatio-spectral interference trace (a) and spatio-temporal intensity reconstruction (b) of a convergent wave (experimental, 35-fs pulses). The amplitude of the plots is in linear scale (see color scale on right).

**D. Spatio-temporal interference of two plane waves**

The spatio-temporal interference of two ultrashort waves constitutes a more complex situation. To create the test beam, we formed a double-beam structure using a Mach-Zehnder interferometer before the test beam input arm of the fiber coupler. Both beams were first aligned and temporally overlapped. Then, we slightly crossed one beam with respect to the other, thus obtaining the spatial interferences of two crossing plane waves. In Fig. 5, we show the experimental results and simulations for this case using a different laser system of 120-fs pulses, also enabling us to reconstruct pulses with a narrower spectrum (FWHM~9 nm). In this case, we used the GRENOUILLE technique to characterize the reference that preserves the spatial homogeneity (required by GRENOUILLE), because we split the laser beam before the Mach-Zehnder device. In Fig. 5a shows the interference spectrum trace, which displays two different fringe patterns: first, the fringes in the spectral dimension corresponding to the spectral interferences between the test beam and a 2.0-ps delayed reference beam, and second the fringes in the spatial dimension (thirteen maxima and minima) arising from the spatial interference of the two crossing waves that form part of the test beam. In this case, we scanned 5 mm of the beam profile in $20\mu m$-steps (251 points). In the spatio-temporal intensity reconstruction (Fig. 5b) the two waves of the test beam had a delay of around zero and had slightly crossing pulse-fronts (relative tilt 36 fs for the 5 mm profile). The maxima and minima of the double wave reconstruction are due to the spatial interference of the beams. We performed the simulations using parameters (spectrum, angle and delay) extracted from the experimental conditions. The interference trace is shown in Fig. 5c, and the intensity reconstruction in Fig. 5d. The simulations and the experiments are in good agreement, showing the same behavior.

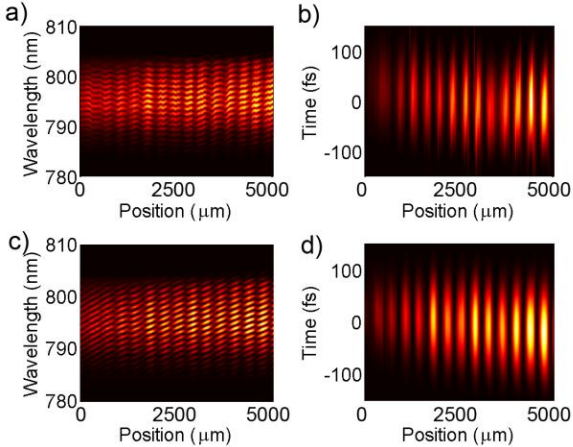

Fig. 5. (Color online) Experimental (exp) and simulated (sim) spatio-spectral interference trace (a: exp, c: sim) and spatio-temporal intensity reconstruction (b: exp, d: sim) of the interference between two crossing waves for 120-fs pulses.

We also implemented the previous experiment with the 35-fs-pulse duration laser, obtaining interferences in a spectral bandwidth of 70 nm. We created the double beam with the Mach-Zehnder interferometer and controlled the relative angle and delay between the beams. In this case the delay between the test and the reference beam was 2.0 ps. We then scanned $10000\mu m$ on a transverse axis of the beam in $20\mu m$-steps (501 points). The experimental results and the corresponding simulations are shown in Fig. 6. The spatially-resolved interference spectrum in Fig. 6a clearly shows the spectral interferences with the reference beam and the spatial interferences of the double wave forming the test pulse. Reconstruction of the spatio-temporal intensity (Fig. 6b) reveals two relatively crossed plane waves. The intensity has the characteristic structure of maxima and minima due to the spatial interferences of the two beams. In this experiment, the angle between the beams was sufficiently high to have 100-fs-separated double pulses on both sides of the beam. In Fig. 6c, we show the temporal profile of the double pulse corresponding to position $7040\mu m$ colored with the instantaneous wavelength and compare with the simulated profile (dashed line), checking that there is no important chirp. Finally, we show the simulated intensity (Fig. 6d) with the parameters involved in the experiment. The simulations match the experimental reconstruction very well.

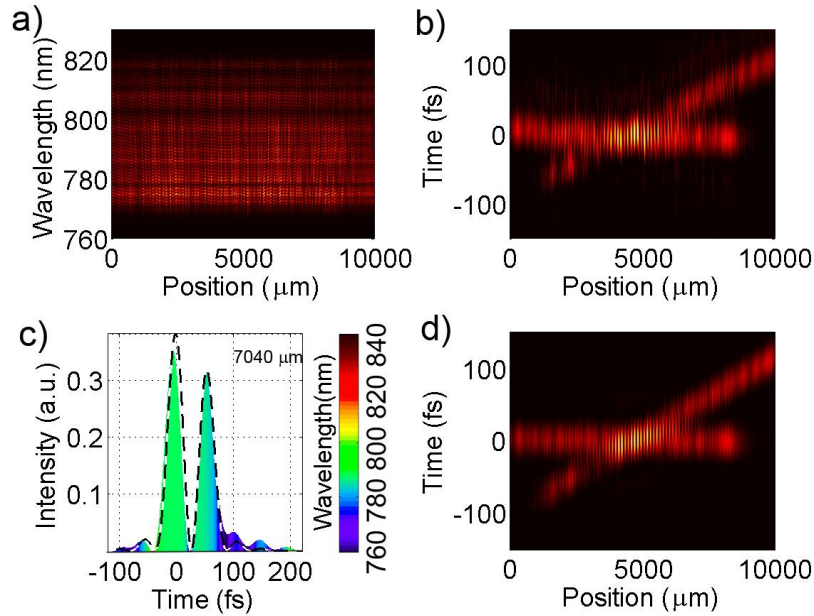

Fig. 6. (Color online) Experimental spatio-spectral interference trace (a) and spatio-temporal intensity reconstruction (b: experimental, d: simulation) of the interference of two crossing waves for 35-fs pulses. Experimental temporal profile and instantaneous wavelength for the 7040 μm position in comparison with the simulated data (dashed line) (d).

Alonso *et al*.

### E. Spatio-temporal interference of a plane and a spherical wave

Finally, we measured the spatio-temporal interference of a spherical and a plane wave structure (Fig. 7). In this experiment, we used 35-fs laser pulses. We used a 50-cm focal lens in one arm of the Mach-Zehnder interferometer to obtain the spherical beam, whereas the other arm controlled the delay between the spherical and the plane wave. The delay between the test and the reference beam was 600 fs, whereas the spherical and the plane wave overlapped in the central region. The spatio-spectral interference pattern of the spherical and plane waves can be seen in Fig. 7a (the test beam trace, without reference). We show the experimental interference spectral trace of the test and reference beams in Fig. 7b, where the quadratic variation of the spectral fringes position due to the curvature of the spherical beam contribution (convergent) can be seen. This trace is the same as the test beam spectrum, the only difference lying in the spectral fringes with the delayed reference pulse. The transverse scan of 4 mm was performed with $8 \mu m$-steps (501 points). The spatio-temporal intensity reconstruction (Fig. 7c) shows the interference of the spherical and the plane wave: a modulated convergent beam is retrieved with maxima and minima in the profile. The spacing of this modulation is larger in the central than in the peripheral region, as corresponds to spherical and plane wave interference, and the same pattern was obtained in the simulation (Fig. 7d). The relative delay between the spherical and the plane beam was zero. We repeated this measurement for different relative delays between the plane and the spherical wave up to 100 fs (above and below), such that in the reconstruction we see how both beams separate in time and the spatial interferences decrease. We also tested this situation with higher delays between the test and the reference beam (1.0, 1.5 and 2.0 ps) and obtained the same intensity reconstructions.

As discussed above, here we demonstrate the stability and reliability of the fiber coupler interferometer. The fluctuations in the experimental measurements and reconstructions in comparison with the simulations are not due to STARFISH limitations but to the laser beam shot-to-shot variations and the beam profile inhomogeneity, which are a consequence of using a terawatt laser with an amplification stage. The acquisition of a full spectral trace usually takes about one minute, such that these variations of the laser pulses may introduce some noise into the reconstructions. Moreover, the use of a Mach-Zehnder interferometer for the double-beam structure has inherent instability and may cause fluctuations in the spatio-temporal interference experiments.

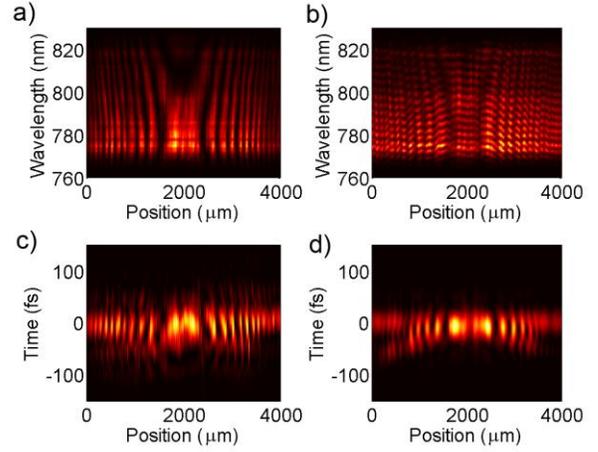

Fig. 7. (Color online) Spherical and plane wave interference for 35-fs pulses. Experimental spatio-spectral test beam trace (a) and interference trace (b). Experimental (c) and simulated (d) spatio-temporal intensity reconstruction.

## 4. CONCLUSIONS

We propose a novel scheme for the spatio-temporal characterization of ultra-short laser pulses based on a fiber-optic coupler interferometer (STARFISH). The device has the advantages of being alignment-free, the fact that only one reference is used, and simplicity; only a fiber coupler and a standard spectrometer are necessary. The time direction is determined with this technique whenever it is known for the reference characterization, as in our case.

We have shown it to measure 1.3-ps-long ($1/e^2$ width) negatively chirped pulses. According to the simulations, it could measure 1-ps FWHM unchirped pulses or, in the case of broadband chirped-pulses, even longer ones. Moreover, the use of better resolution spectrometers allows the measurement of much longer pulses, conserving the advantages of the fiber coupler since it is only necessary to connect the fiber output to the spectrometer entrance. We applied STARFISH to the characterization of a converging beam but also to more complex structures, such as measurement of the spatio-temporal interference of plane-plane and spherical-plane waves, the results obtained being in agreement with the simulations. We reconstructed laser beams using two laser systems with different pulse durations (35 and 120 fs) and spectral bandwidths. Despite working with terawatt lasers at 10 Hz, which do not have a perfectly homogeneous profile and are unstable, we demonstrate the ability of our method to reconstruct complex pulses.

We expect this system to be used to characterize laser beams after they have undergone certain non-linear processes or have passed through certain optical systems.

## ACKNOWLEDGMENTS

The authors are grateful to S. Jarabo from the University of Zaragoza (Spain) for his assistance regarding the fiber coupler. We acknowledge support from Spanish Ministerio de Ciencia e Innovación (MICINN) through the Consolider Program SAUUL (CSD2007-00013), Research project FIS2009-09522 and from the Junta de Castilla y León through the Program for Groups of Excellence (GR27). We also acknowledge support from the Centro de Laseres Pulsados, CLPU, Salamanca, Spain. B. Alonso and I. J. Sola acknowledge support from the MICINN through the FPU and "Ramón y Cajal" programs respectively.